\documentclass[a4paper,twocolumn]{article}
\usepackage{graphicx}
\usepackage{amsmath}

\begin{document}

\title{The entanglement of some non-two-colorable graph states}
\author{Xiao-yu Chen , Li-zhen Jiang  \\
\small{College of Information and Electronic Engineering, Zhejiang
Gongshang University, Hangzhou 310018,People's Republic of China
}}
\date{}
\maketitle

\begin{abstract}
We exactly evaluate the entanglement of a six vertex and a nine vertex graph
states which correspond to non ''two-colorable'' graphs. The upper bound of
entanglement for five vertices ring graph state is improved to 2.9275, less
than upper bound determined by LOCC. An upper bound of entanglement is
proposed based on the definition of graph state.

PACS number(s): 03.67.Mn, 03.65.Ud,

Keyword(s):graph state; closest separable state; multipartite entanglement
\end{abstract}


\section{Introduction}

Entanglement is one of the most important concepts and resources in quantum
information theory. However, the quantification of the entanglement of a
given quantum state is quite difficult except for bipartite pure state,
where the distance-like measure of the entanglement (\textit{Relative
Entropy of Entanglement})\cite{Vedral1} \cite{Vedral2} and the operational
measures of entanglement (\textit{\ Entanglement of Formation, Distillable
Entanglement }) \cite{Bennett}are all equal to the entropy of the reduced
state obtained by tracing out one part of the pure state. In bipartite
system, apart from pure state, the entanglement of a mixed state is not easy
to calculate in general if not impossible. The situation becomes even worse
for multipartite system where the basic states corresponding to Bell basis
are not clearly recognized \cite{Dur}\cite{Miyake}. Thus the extension of
operational entanglement measures to multipartite system is not available.
Nevertheless, a variety of different entanglement measure have been proposed
for multipartite setting. Among them are the \textit{(Global) Robustness of
Entanglement }\cite{Vidal}\textit{\ , }the \textit{Relative Entropy of
Entanglement, }and \textit{the Geometric Measure }\cite{Wei1}\textit{. } The
robustness measures the minimal noise (arbitrary state) that we need to
added to make the state separable. The geometric measure is the distance of
state to the closest product state in terms of the fidelity. The relative
entropy of entanglement is a valid entanglement measure for multipartite
state, it is the relative entropy of the state under consideration to the
closest fully separable state.

The quantification of multipartite entanglement is usually very difficult as
most measures are defined as the solutions to difficult variational
problems. Even for pure multipartite state, the entanglement can only be
obtained for some special scenario. Fortunately, due to the inequality on
the logarithmic robustness, relative entropy of entanglement and geometric
measure of entanglement\cite{Wei2} \cite{Hayashi1} \cite{Wei3}, these
entanglement measures are all equal for stabilizer states \cite{Hayashi2} .
Thus for stabilizer state $\left| S\right\rangle $, the entanglement can be
written as
\begin{equation}
E=\min_\phi -\log _2\left| \left\langle S\right| \left. \phi \right\rangle
\right| ^2,
\end{equation}
where $\phi =\bigotimes_j(\sqrt{p_j}\left| 0\right\rangle +\sqrt{1-p_j}%
e^{i\varphi _j}\left| 1\right\rangle )$ is the separable pure state.

The entanglement is upper bounded by the local operation and classical
communication (LOCC) bound $E_{LOCC}$ , and lower bounded by some bipartite
entanglement deduced from the state, that is, the 'matching' bound $E_{bi}$
\cite{Markham}. It is well known that all graph states are stabilizer
states, so the inequality for the entanglement of a graph state is
\begin{equation}
E_{bi}\leq E\leq E_{LOCC}.
\end{equation}
If the lower bound coincides with the upper bound, the the entanglement of
the graph state can be obtained. This is the case for 'two-colorable' graph
states such as multipartite GHZ states, Steane code, cluster state, and
state of ring graph with even vertices. For a state of ring graph with odd $%
n $ vertices, we have $\left\lfloor \frac n2\right\rfloor \leq E\leq
\left\lceil \frac n2\right\rceil $ \cite{Markham}$.$

We in this paper will concern with the entanglement of a graph state whose
graph is not ''two-colorable''. A new upper bound based directly on the
definition of graph state is proposed. The symmetry of the graph is utilized
to further reduce the upper bounds for some highly symmetric graph states,
including the five vertices ring and Peterson graph.

\section{Graph state}

A graph $G=(V;\Gamma )$ is composed of a set $V$ of $n$ vertices and a set
of edges specified by the adjacency matrix $\Gamma $, which is an $n\times n$
symmetric matrix with vanishing diagonal entries and $\Gamma _{ab}$ $=1$ if
vertices $a,b$ are connected and $\Gamma _{ab}$ $=0$ otherwise. The
neighborhood of a vertex $a$ is denoted by $N_a$ $=\{v\in V\left| \Gamma
_{av}=1\right. \}$, i.e, the set of all the vertices that are connected to $%
a $. Graph states \cite{Hein1} \cite{Sch}] are useful multipartite entangled
states that are essential resources for the one-way computing \cite{Raus}
and can be experimentally demonstrated \cite{Walther}\cite{Lu}. To associate
the graph state to the underlying graph, we assign each vertex with a qubit,
each edge represents the interaction between the corresponding two qubits.
More physically, the interaction may be Ising interaction of spin qubits.
Let us denote the Pauli matrices at the qubit $a$ by $X_a,Y_a,Z_a$ and
identity by $I_a$. The graph state related to graph $G$ is defined as
\begin{equation}
\left| G\right\rangle =\prod_{\Gamma _{ab}=1}U_{ab}\left| +\right\rangle
_x^V=\frac 1{\sqrt{2^n}}\sum_{\mathbf{\mu }=\mathbf{0}}^{\mathbf{1}%
}(-1)^{\frac 12\mathbf{\mu }\Gamma \mathbf{\mu }^T}\left| \mathbf{\mu }%
\right\rangle _z  \label{wave0}
\end{equation}
where $\left| \mathbf{\mu }\right\rangle _z$ is the joint eigenstate of
Pauli operators $Z_a$ ($a\in V$) with eigenvalues $(-1)^{\mu _a}$, $\left|
+\right\rangle _x^V$ is the joint +1 eigenstate of Pauli operators $X_a$ ( $%
a\in V$) , and $U_{ab}$ ($U_{ab}=diag\{1,1,1,-1\}$ in the $Z$ basis) is the
controlled phase gate between qubits $a$ and $b$. Graph state can also be
viewed as the result of successively performing 2-qubit Control-Z operations
$U_{ab}$ to the initially unconnected $n$ qubit state $\left| +\right\rangle
_x^V$. It can be shown that graph state is the joint $+1$ eigenstate of the $%
n$ vertices stabilizers
\begin{equation}
K_a=X_a\prod_{b\in N_a}Z_b:=X_aZ_{N_a},\text{ }a\in V.
\end{equation}
Meanwhile, the graph state basis are $\left| G_{k_1,k_2,\cdots
k_n}\right\rangle $ $=\prod_{a\in V}Z_a^{k_a}\left| G\right\rangle ,$ with $%
k_a=0,1.$ Since all of the graph basis states are local unitary equivalent,
they all have equal entanglement, so we only need to determine the
entanglement of graph state $\left| G\right\rangle $. Once the entanglement
of a graph state is obtained, the entanglement of all the graph basis states
are obtained.

\section{The upper bound of graph state}

The fidelity $F_\phi =\left| \left\langle G\right| \left. \phi \right\rangle
\right| ^2$plays a crucial rule in calculating the entanglement. For a graph
state, we have
\begin{equation}
E=\min_\phi -\log _2\left| \left\langle G\right| \left. \phi \right\rangle
\right| ^2=-\log _2(\max_\phi F_\phi ).
\end{equation}
Denote $F=\max_\phi F_\phi $ as the fidelity between the graph state and the
closest pure separable state. One of the ways to obtain the upper bound the
entanglement is to relax the maximization. For two-colorable graph, the set
with majority vertices is colored with Amber, the set with minority vertices
is colored with Blue. Without loss of generality, the Amber colored vertices
are labelled as $a=1,\ldots ,\left| A\right| ,$ the Blue vertices are
labelled as $b=\left| A\right| +1,\ldots ,n$. Since all Amber vertices are
not adjacent with each other, we can perform $X_a$ ($a=1,\ldots ,\left|
A\right| $) measurements to all Amber qubits simultaneously. And applying $%
Z_b$ ($b=\left| A\right| +1,\ldots ,n$) measurements to all Blue qubits at
the same time. Thus all Amber stabilizers $K_a$ can be measured
simultaneously by LOCC. The maximal number of states that can be
discriminated by LOCC then is $2^{\left| A\right| }$ according to the theory
of graph state basis \cite{Markham}. Applying the inequality on the
relationship of LOCC discrimination of states and the entanglement\cite
{Hayashi1}, one has $\left| A\right| \leq n-E,$ that is,
\begin{equation}
E\leq n-\left| A\right| .  \label{wave1}
\end{equation}
This upper bound of LOCC may be extended to graphs that are not
two-colorable by some modification. However, it is possible to obtain the
upper bound without the LOCC state discrimination.

We will obtain the upper bound of the entanglement with the definition of
the graph state. The graph state may not be two-colorable. Suppose the
maximal non-adjacent vertices set $A$ has $\left| A\right| $ vertices. As
before, we label these vertices with $a=1,\ldots ,\left| A\right| $. The
other vertices are in the set $B=V-A,$ the vertices are labelled with $%
b=\left| A\right| +1,\ldots ,n.$ Note that the vertices within set $B$ may
connect with each other, for the graph may not be two-colorable. The
adjacency matrix $\Gamma $ now is
\begin{equation}
\Gamma =\left[
\begin{array}{ll}
\Gamma _A & \Gamma _{AB} \\
\Gamma _{AB}^T & \Gamma _B
\end{array}
\right] .
\end{equation}
Since any vertices pairs are not adjacent in set $A$, the adjacency matrix
of set $A$ is an all zero $\left| A\right| \times \left| A\right| $ matrix,
\begin{equation}
\Gamma _A=\mathbf{0.}
\end{equation}
Denote $\mathbf{\mu =(\mu }_A,\mathbf{\mu }_B\mathbf{),}$ where the binary
vectors $\mathbf{\mu }_A=(\mu _1,\ldots ,\mu _{\left| A\right| }),$ $\mathbf{%
\mu }_B=(\mu _{\left| A\right| +1},\ldots ,\mu _n),$ then the graph state
can be written as $\left| G\right\rangle =$ $\left| G_1\right\rangle +$ $%
\left| G_2\right\rangle $, the unnormalized states (in $Z$ basis )
\begin{eqnarray}
\left| G_1\right\rangle &=&\frac 1{\sqrt{2^n}}\sum_{\mathbf{\mu }_A=\mathbf{0%
}}^{\mathbf{1}}(-1)^{\frac 12(\mathbf{\mu }_A,\mathbf{0)}\Gamma (\mathbf{\mu
}_A,\mathbf{0)}^T}\left| \mathbf{\mu }_A,\mathbf{0}\right\rangle  \nonumber
\\
&=&\frac 1{\sqrt{2^n}}\sum_{\mathbf{\mu }_A=\mathbf{0}}^{\mathbf{1}}\left|
\mathbf{\mu }_A,\mathbf{0}\right\rangle ,
\end{eqnarray}
where we have used the fact that
\begin{equation}
\frac 12(\mathbf{\mu }_A,\mathbf{0)}\left[
\begin{array}{ll}
\mathbf{0} & \Gamma _{AB} \\
\Gamma _{AB}^T & \Gamma _B
\end{array}
\right] (\mathbf{\mu }_A,\mathbf{0)}^T=0.
\end{equation}
And
\begin{equation}
\left| G_2\right\rangle =\frac 1{\sqrt{2^n}}\sum_{\mathbf{\mu }_A=\mathbf{0}%
}^{\mathbf{1}}\sum_{\mathbf{\mu }_B\neq \mathbf{0}}(-1)^{\frac 12(\mathbf{%
\mu }_A,\mathbf{\mu }_B\mathbf{)}\Gamma (\mathbf{\mu }_A,\mathbf{\mu }_B%
\mathbf{)}^T}\left| \mathbf{\mu }_A,\mathbf{\mu }_B\right\rangle
\end{equation}
To obtain a lower bound of the extremal fidelity $F$, we can choose
\begin{equation}
\left| \phi \right\rangle =\bigotimes_{a=1}^{\left| A\right| }(\sqrt{p_a}%
\left| 0\right\rangle +\sqrt{1-p_a}e^{i\varphi _a}\left| 1\right\rangle
)\otimes \left| 0\right\rangle ^{\otimes (n-\left| A\right| )}.
\label{wave2}
\end{equation}
Since $\mathbf{\mu }_B\neq \mathbf{0}$ in the state $\left| G_2\right\rangle
$ and the last $(n-\left| A\right| )$ qubits of $\left| \phi \right\rangle $
are all in $\left| 0\right\rangle ,$ we have $\left\langle G_2\right| \left.
\phi \right\rangle =0.$ Thus one has
\begin{eqnarray}
\left\langle G\right| \left. \phi \right\rangle &=&\left\langle G_1\right|
\left. \phi \right\rangle =\frac 1{\sqrt{2^n}}\sum_{\mathbf{\mu }_A=\mathbf{0%
}}^{\mathbf{1}}\left\langle \mathbf{\mu }_A\right| \bigotimes_{a=1}^{\left|
A\right| }(\sqrt{p_a}\left| 0\right\rangle  \nonumber \\
&&+\sqrt{1-p_a}e^{i\varphi _a}\left| 1\right\rangle )  \nonumber \\
&=&\frac 1{\sqrt{2^n}}\prod_{a=1}^{\left| A\right| }(\sqrt{p_a}+\sqrt{1-p_a}%
e^{i\varphi _a}).
\end{eqnarray}
The maximal fidelity for separable state (\ref{wave2}) is
\begin{equation}
F_0=\max_{p_a,\varphi _a}\left| \left\langle G\right| \left. \phi
\right\rangle \right| ^2=2^{-(n-\left| A\right| )},
\end{equation}
which is achieved when $p_a=\frac 12,$ $\varphi _a=0,\pi $ ($a=1,\ldots
,\left| A\right| $). The upper bound of the entanglement is
\begin{equation}
-\log _2F_0=n-\left| A\right| ,
\end{equation}
where $\left| A\right| $ is the maximal number of non-adjacent vertices of
the graph. Note that we obtain the upper bound without the knowledge of
LOCC, and the result can be applied to any graph state.

As an application, we consider the graph in Fig.1 (a). The stabilizer code
based on this graph is [[6,1,3]] quantum error correction code. The maximal
number of non-adjacent vertices of the graph is $3$ (i.e. vertices $1,3,6$
or vertices $2,4,6$). Thus we have the entanglement upper bound $6-3=3.$
Meanwhile, the lower 'matching' bound \cite{Markham} is also $3.$ This is
due to the fact that multipartite entanglement is no less than the
corresponding bipartite entanglement \cite{Markham}. We have $E\geq E_{bi}.$
The lower bound $E_{bi}$ is obtained with a bipartition of the graph into
subgraph $C=\{V_C,\Gamma _C\}$ and $D=\{V_D,\Gamma _D\},$with $V_C=\{1,4,5\}$
and $V_D=\{2,3,6\}.$ Removing the local edges in both parties by local
Control-Z operations, we obtain $3$ Bell pairs between the two parties. So $%
E_{bi}=3.$ The detail process is to apply $U_{15}$ and $U_{45}$ operations
which are local in subgraph $C$, and apply $U_{23}$ operation which is local
in subgraph $D.$ The edges of the remain graph are $(1,2),(3,4)$ and $(5,6).$
The upper bound and the lower bound coincide, thus the entanglement of the
graph state is $3$.

\begin{figure}[tbp]
\includegraphics[ trim=0.000000in 0.000000in -0.138042in 0.000000in,
height=2.0in, width=3.5in ]{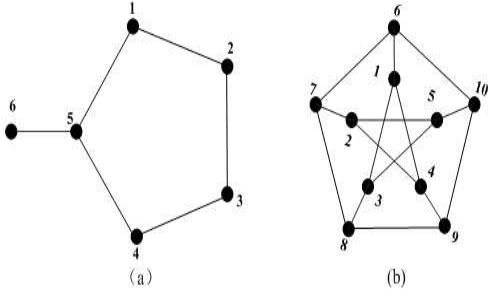}
\caption{(a) The graph for [[6,1,3]] code; (b) Peterson graph.}
\end{figure}

\section{Improving the upper bound with symmetry}

For a graph state of a ring graph with odd $n$ vertices, we have $%
\left\lfloor \frac n2\right\rfloor \leq E\leq \left\lceil \frac
n2\right\rceil .$ The upper bound is obtained by LOCC and also by our
non-adjacent vertices set method. We will show that this upper bound can be
further improved for five vertices ring graph state by making use of the
symmetry of the graph. Our symmetrical consideration also shows that the
Peterson graph \cite{Hein2} state has entanglement upper bound of $5$, while
LOCC and non-adjacent vertices set method can only give the upper bound of $%
6 $.

\subsection{Five vertices ring}

The graph of five vertices ring is the underlying graph of the famous
[[5,1,3]] stabilizer code. Denote the codewords of [[5,1,3]] as $\left|
\overline{0}\right\rangle $ and $\left| \overline{1}\right\rangle $\cite
{Gottesman}$,$ the graph state will be $\left| G\right\rangle =\frac 1{\sqrt{%
2}}(\left| \overline{0}\right\rangle $ $-\left| \overline{1}\right\rangle ).$
We simply suppose the separable state be
\begin{equation}
\left| \phi \right\rangle =(\sqrt{p}\left| 0\right\rangle +\sqrt{1-p}%
e^{i\varphi }\left| 1\right\rangle )^{\otimes 5},
\end{equation}
where we have considered the symmetry of the graph. Denote $x=\sqrt{p},y=%
\sqrt{1-p}e^{i\varphi },$ the fidelity $F_G=\left| \left\langle G\right|
\left. \phi \right\rangle \right| ^2$ will be
\begin{equation}
F_G=\frac 1{32}\left| x^5-y^5+5x^4y-5xy^4\right| ^2.  \label{wave3}
\end{equation}
With a numerical calculation, the entanglement upper bound can be found to
be
\begin{equation}
\min_{p,\varphi }-\log _2F_G\approx 2.9275,
\end{equation}

A more precise condition for maximum of $F_G$ can be obtained. Let us
consider the fidelity $F_{\overline{0}}=\left| \left\langle \overline{0}%
\right| \left. \phi \right\rangle \right| ^2,$%
\begin{equation}
F_{\overline{0}}=\frac 1{16}\left| x^5-5xy^4\right| ^2.
\end{equation}
The fidelity $F_{\overline{0}}$ can be rewritten as $F_{\overline{0}}=\frac
1{16}(p^5+25p(1-p)^4-10p^3(1-p)^2\cos 4\varphi ).$ The maximal will be
achieved when $\cos 4\varphi =-1,$ so $F_{\overline{0}}=\frac
p{16}[p^2+5(1-p)^2]^2.$ The derivative $\frac{dF_{\overline{0}}}{dp}=0$
reduces to $6p^2-6p+1=0,$ which is $p=$ $\frac 12(1\pm \frac 1{\sqrt{3}}).$
For $p=$ $\frac 12(1-\frac 1{\sqrt{3}}),$ $\frac{d^2F_{\overline{0}}}{dp^2}%
=-\frac 54(1+\sqrt{3})<0;$ For $p=$ $\frac 12(1+\frac 1{\sqrt{3}}),$ $\frac{%
d^2F_{\overline{0}}}{dp^2}=\frac 54(\sqrt{3}-1)>0.$ Thus the fidelity
reaches its maximal $F_{\overline{0}\max }=\frac{3+\sqrt{3}}{36}$ when $%
\varphi =\pm \frac \pi 4,\pm \frac{3\pi }4,$ and $p=\frac 12(1-\frac 1{\sqrt{%
3}}).$ It is interesting that when at $p=\frac 12(1-\frac 1{\sqrt{3}}),$ $%
\varphi =\pm \frac \pi 4,$ we have the the maximal fidelity $F_{G\max }=%
\frac{3+\sqrt{3}}{36}.$ We may calculate the derivatives of $F_G$ at points $%
(p,\varphi )=(\frac 12(1-\frac 1{\sqrt{3}}),\pm \frac \pi 4),$ the first
order derivatives are $\frac{\partial F_G}{\partial p}=0,\frac{\partial F_G}{%
\partial \varphi }=0,$ the second order derivatives are $\frac{\partial ^2F_G%
}{\partial p^2}=-\frac 54<0,$ $\frac{\partial ^2F_G}{\partial p\partial
\varphi }=\pm \frac 5{12},$ $\frac{\partial ^2F_G}{\partial \varphi ^2}%
=-\frac 5{108}(3+2\sqrt{3})<0.$ The Jacobian is
\begin{equation}
J=\left|
\begin{array}{ll}
\frac{\partial ^2F_G}{\partial p^2}, & \frac{\partial ^2F_G}{\partial
p\partial \varphi } \\
\frac{\partial ^2F_G}{\partial p\partial \varphi } & \frac{\partial ^2F_G}{%
\partial \varphi ^2}
\end{array}
\right| =\frac{25\sqrt{3}}{216}>0.
\end{equation}
Thus $(p,\varphi )=(\frac 12(1-\frac 1{\sqrt{3}}),\pm \frac \pi 4)\ $are the
points of maximal fidelity $F_G.$ We can also prove that $(p,\varphi
)=(\frac 12(1+\frac 1{\sqrt{3}}),\pm \frac \pi 4)\ $are the points of
maximal fidelity $F_G.$

$.$

The entanglement upper bound is
\begin{equation}
\min_{p,\varphi }-\log _2F_G=-\log _2\frac{3+\sqrt{3}}{36}\approx 2.9275.
\end{equation}
It is less than $3,$ the best upper bound by LOCC or non-adjacent vertices
set method.

Although we use the identical product state to obtain the upper bound of the
entanglement, and this upper bound is far from the lower bound which is $2$,
a random search calculation indicates that this upper bound is possibly the
entanglement itself.

\subsection{Peterson graph}

For Peterson graph $G_P$ in Fig. 1 (b), the lower bipartite bound for the
entanglement of the graph state is easily obtained to be $5$, which is the
number of Bell pairs between the subgraph $C$ with $V_C=\{1,2,3,4,5\}$ and
subgraph $D$ with $V_D=\{6,7,8,9,10\}.$ The number of maximal non-adjacent
vertices set is $4$, thus the entanglement upper bound is $10-4=6.$

Suppose the separable state be
\begin{equation}
\left| \phi \right\rangle =(\sqrt{p}\left| 0\right\rangle +\sqrt{1-p}%
e^{i\varphi }\left| 1\right\rangle )^{\otimes 10}.
\end{equation}
Denote $x=\sqrt{p},y=\sqrt{1-p}e^{i\varphi }.$ We have
\begin{equation}
\left\langle G_P\right| \left. \phi \right\rangle
=2^{-5}\sum_{j=0}^{10}c_jx^{10-j}y^j.
\end{equation}
The coefficient $c_j=\sum_{\mathbf{\mu }\in \Lambda _j}(-1)^{\frac 12\mathbf{%
\mu }\Gamma \mathbf{\mu }^T}$, where $\Lambda _j=\{\mathbf{\mu }$ $\left|
\sum_{k=1}^{10}\mu _k=j\right\} $. The coefficient vector is
\begin{equation}
\mathbf{c=(}1,10,15,0,-50,108,50,0,-15,10,-1).
\end{equation}
A rather special closest separable state is with $p=\frac 12,$ $\varphi
=\frac \pi 2.$ The maximal fidelity is
\begin{equation}
F=\left| 2^{-3}(-1+i)\right| ^2=\frac 1{32}.
\end{equation}
The entanglement upper bound coincides with its lower bound. The
entanglement of the Peterson graph state is $5.$

\section{Conclusions}

We have proposed an upper bound for the entanglement of a graph state. The
bound is based on the definition of the graph state. We obtain the bound by
calculating the fidelity of the graph state with respect to some separable
state. The vertices of the graph are divided into two subsets, one with all
its vertices that are not adjacent with each other, and we make this subset
as large as possible and it has $\left| A\right| $ vertices. Then the
entanglement of the $n$ vertices graph state is upper bounded by $n-\left|
A\right| .$ The entanglement measure can be the \textit{(Global) Robustness
of Entanglement, }the \textit{Relative Entropy of Entanglement, }and \textit{%
the Geometric Measure. }These measures are all equal for graph states. Using
this bound, we find the entanglement of graph state which [[6,1,3]] code
based on to be $3.$ The upper bound of the graph state has been further
improved for some highly symmetric states. These states are five vertices
ring graph state and Peterson graph state. With the product of identical
qubit states, we find that the entanglement upper bound for five vertices
ring graph state is about $2.9275,$ which is less than $3$, the bound given
by LOCC and our non-adjacent vertices set method. We also determine the
entanglement of Peterson graph state to be $5$ (less than $6$ given by LOCC)
by using the product of identical qubit states as the closest separable
state.

\end{document}